\documentclass{iopart}
\usepackage[T1]{fontenc}
\usepackage[latin1]{inputenc}
\usepackage{verbatim}
\usepackage{hyperref}
\usepackage{iopams}
\usepackage{graphicx}

\begin{document}

\title{Protein Simulations combining an All-Atom Force Field with a Go-Term}

\author{Jan H Meinke$^1$ and Ulrich H E Hansmann$^{1,2}$}

\address{$^1$ John-von-Neumann Institute for Computing, Forschungszentrum J\"{u}lich,
D-52425 J\"{u}lich, Germany}

\address{$^2$ Department of Physics, Michigan Technological University, Houghton, MI 49931, USA}

\eads{\mailto{j.meinke@fz-juelich.de}, \mailto{u.hansmann@fz-juelich.de}}

\begin{abstract}
Using a variant of parallel tempering, we study the changes in sampling within a simulation 
when the all-atom model is coupled to a Go-like potential. We find that the native structure
is not the lowest-energy configuration in the all-atom force field. Adding a Go-term deforms 
the energy landscape in a way that the native configuration becomes the global minimum.
\end{abstract}



\section{Introduction}
Most proteins  exist  at room temperature in a {\em unique} structure that 
one can  identify with the lowest {\em potential} energy 
conformation \cite{Anfinsen1973}.  It is now commonly assumed that the energy 
landscape of a protein is shaped like a funnel with the native state at the 
bottom \cite{Onuchic1997}. At the same time, the landscape has many deep local
minima and high barriers. This is because the  average protein in the 
contains thousands of atoms, and interactions between the atoms can be both 
repulsive and attractive.

Due to the large number of continuous degrees of freedom and the rough
energy landscape simulating proteins remains a computational challenge.
The time to find the native structure of a protein (the bottom of the funnel)
 depends both on the roughness of the energy landscape and the steepness
 of the funnel. The more pronounced the funnel is, the faster the protein will fold.
 This is one reason for the popularity of the Go model \cite{Abe81, Go1983, Takeda1999}.
 Its basic assumption is that only interaction present in the native state of a protein are relevant for the folding process. An appropriate energy function then ignores non-native interactions and rewards native interactions. Hence, the Go model  represents a perfect funnel model and has none of the roughness normally associated with the protein-folding energy landscape. In their 1981 paper \cite{Abe81},e.g, Abe and Go used a lattice model where each amino acid occupied a single lattice site. If two amino acids are on adjacent sites that are neighbors in the native state, the system gained $\epsilon$ in energy. This type of contact potential inherently cannot distinguish between the original native state and its mirror image.
%
 Go-like energy terms are usually only defined between heavy atoms in 
 the protein backbone
%
 and therefore lack the detail of all-atom force fields. 
 On the other hand, all-atom simulations relying on present energy functions 
utilize a number of approximations that may lead to additional spurious minima \cite{Hansmann2004,Hansmann2003,Trebst05}
and therefore to an energy landscape with an artificially increased roughness. As a consequence,
all-atom simulations are usually too slow to allow an efficient study of the folding of stable
domains in proteins, which contain of the order of 50--200 residues). 

To speed up all-atom simulations one could deform the energy landscape to obtain a
steeper folding funnel. In principle, this can be done by adding a Go-like term to the all-atom
energy function. For instance,  Pogorelov and Luthey-Schulten used this method to speed up 
molecular dynamics simulations of the folding of the $\lambda$-repressor \cite{Pogorelov2004}.
It is not clear, however, what the optimal coupling is, how the speed up depends on the coupling, and at what coupling the system is dominated by the Go-term.

\begin{figure}
\centering
\includegraphics[width=1\columnwidth,height=0.25\textheight,keepaspectratio]{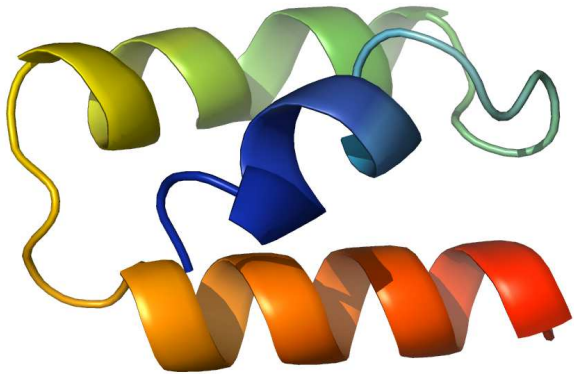}
\caption{\label{fig:Protein-A}Native structure of the 46 amino-acid long
segment of Protein A (1bdd) used as native reference structure. The
structure taken from the Protein Data Bank has been adjusted to fit
the standard geometry assumed by ECEPP/3, where the bond lengths are
fixed. The ground state consists of three helices and two loops connecting
the helices. }
\end{figure}

We have studied these questions using  a 46 residue segment of Protein A (1bdd
in the Protein Data Bank) and a variant of the parallel tempering method that will 
be introduced in the next chapter.   The structure of the protein is shown 
 in \fref{fig:Protein-A}. The
segment consists of 3 helices and short loops connecting the helices
. %
In the following we will first introduce our method followed by our results and concluding remarks.

\section{Methods}

Our investigations rely on simulations of Protein A with the ECEPP/3 force field \cite{ECEPP2, ECEPP3}. This force field is implemented in the 2005 version of the program package SMMP \cite{Eisenmenger2001,Eisenmenger2006}.
 The interactions between the atoms within a protein are approximated
 by a sum $E_\mathrm{ECEPP/3}$ consisting of electrostatic energy $E_\mathrm{C}$, a  Lennard-Jones term $E_\mathrm{LJ}$, 
 a hydrogen-bonding term $E_\mathrm{hb}$, and a torsion energy $E_\mathrm{tor}$:  
 \begin{eqnarray}
  E_{\mathrm{ECEPP/3}} &=& E_\mathrm{C} + E_\mathrm{LJ}  + E_\mathrm{hb} + E_\mathrm{tor} \nonumber \\
  &=&  \sum_{(i,j)} \frac{332 q_i q_j}{\epsilon r_{ij}} \nonumber \\
 &&   + \sum_{(i,j)} \left( \frac{A_{ij}}{r_{ij}^{12}} - \frac{B_{ij}}{r_{ij}^6} \right) \nonumber \\ 
 &&   + \sum_{(i,j)} \left( \frac{C_{ij}}{r_{ij}^{12}} - \frac{D_{ij}}{r_{ij}^{10}} \right) \nonumber \\ 
  && + \sum_l U_l ( 1\pm \cos(n_l \xi_l)) \;,
\label{energy}
\end{eqnarray}
where $r_{ij}$ is the distance between the atoms $i$ and $j$,  $\xi_l$ is the $l$-th torsion 
angle.  
%
The factor 332 converts the electrostatic energy into kcal/mol. The charges $q_i$ are partial charges on the atoms. The factors $A_{ij}$, $B_{ij}$, $C_{ij}$, and $D_{ij}$ depend on the type of atoms involved. The factors $U_l$ depend on the residue and the type of dihedral angle. All of these values have been determined empirically and are given in \cite{ECEPP2,ECEPP3}. The magnitudes are chosen such that energies are measured in kcal/mol. 
The all-atom energy of our molecule is the sum of the intra-molecular interactions and the
ones between protein and the surrounding solvent:
\begin{equation}
  E_\mathrm{aa} =  E_{\mathrm{ECEPP/3}} + E_\mathrm{solv}~,
  \end{equation}
 where the protein-solvent interactions are approximated by a solvent accessible surface term 
\begin{equation}
  E_\mathrm{solv} = \sum_i \sigma_i A_i \;.
\label{solventEnergy}
\end{equation}
The sum goes over the  solvent accessible areas $A_i$ of all atoms $i$ weighted by solvation
parameters $\sigma_i$ as determined in \cite{OONS}, a common choice when the 
ECEPP/3 force field is utilized. Note that $E_\mathrm{solv}$ is a rather crude approximation of the
interaction between the polypeptide and the surrounding water that is motivated by the
low computational costs when compared to simulations with explicit water molecules. 

The competing interactions in this detailed energy function lead  to an
energy landscape that is characterized by a multitude of minima separated by high
energy barriers. As the probability to cross an energy barrier of height $\Delta E$ is given by
$\exp (-\Delta E/k_\mathrm{B}T)$ ($k_\mathrm{B}$ the Boltzmann constant) it follows that  extremely long runs are necessary to obtain sufficient statistics in regular canonical simulations at a low temperature $T$.

One popular method to overcome the
resulting  extremely slow thermalization at low temperatures is
parallel tempering \cite{PT1,PT2} (also known as replica exchange 
method or Multiple Markov chains), a techniques that was first 
applied to protein studies in \cite{Hansmann1997}.
In its most common form,  one considers  in parallel tempering
\index{parallel tempering}
an artificial system built up of {\em N non--interacting} replicas 
of the molecule, each at a different temperature $T_i$.
In addition to standard Monte Carlo or molecular dynamics  moves  that act only
on one replica (i.e., the molecule at a fixed temperature), an exchange
of conformations between two copies $i$ and $j=i+1$ is allowed
with probability
\begin{equation}
\fl w({\bf C}^\mathrm{old} \rightarrow {\bf C}^\mathrm{new}) = \min (1,
\exp(-\beta_i E(C_j) - \beta_j E(C_i) + \beta_i E(C_i) +\beta_j E(C_j)))~.
\end{equation}
The exchange of conformations lead to
a faster convergence of  the Markov chain at low temperatures 
than is observed in regular canonical simulations with
only local moves. This is because the resulting random walk in
temperatures allows the configurations to move out of local minima
and cross energy barriers.

While parallel tempering is traditionally done in temperature space,
it can be used with varying potentials as well. The system could be
coarse grained across replicas, or the solvent terms could be varied.
In this paper we vary in some simulations the strength of an additional Go-like
potential term instead of the temperature. With this we can study the effect of a Go-like potential
on the statistics of a Monte Carlo simulation of a protein. 
Go-like potentials have their origin in lattice models. They reward native
contacts with a reduction in energy. If we assume that long- and short-range
interactions cooperatively fold the protein into its native structure
--- this idea is often depicted as a funnel-like structure of the
energy landscape ---, an additional Go-like potential smoothes the
energy landscape which  should lead to faster folding. 

With the added Go-like energy our energy function becomes
\begin{equation}
E_\mathrm{tot}= E_\mathrm{aa}+k_\mathrm{Go}E_\mathrm{Go},
\end{equation}
where $E_\mathrm{aa}$ is the all-atom energy defined above and $k_\mathrm{Go}$ a parameter
that describes the strength of coupling between the two energies.
We use the same form for the Go-like energy term as Pogorelov and Luthey-Schulten \cite{Pogorelov2004}. It is based on an associative memory Hamiltonian with a single
memory. Associative memory Hamiltonians have been used successfully to recognize tertiary structures in proteins \cite{Friedrichs1989} and to study protein folding \cite{Hardin1999}. Associative memory Hamiltonians capture the long-range effects of protein folding better than, for example, a square well. The form used here can be viewed as a continuum model of the original Go lattice model.
\begin{equation}
%
E_\mathrm{Go}=\sum_{i}^{N_{c_{\alpha}}}\sum_{j\neq i,\pm1,\pm2}^{N_{c_{\alpha}}}\gamma_{ij}\times\exp\left[-\frac{\left(r_{ij}-r_{ij}^{\mathrm{Nat}}\right)^{2}}{\left(|i-j|^{0.15}\right)^{2}}\right].
\label{go-term}
\end{equation}
The value of $\gamma_{ij}$ were chosen as in \cite{Pogorelov2004} as $\gamma_{ij}=0.4$ if $3\le|i=j|<9$ and $\gamma_{ij}=0.5$ if $|i-j|\ge9$, where $i$ and $j$ are the indices of the residues.

We also define an order parameter $Q$
\begin{equation}
 Q=\frac{1}{N_{\mathrm {contacts}}}\sum_{i}^{N_{c_{\alpha}}}\sum_{j\neq i,\pm1}^{N_{c_{\alpha}}}\gamma_{ij}\times\exp\left[-\frac{\left(r_{ij}-r_{ij}^{\mathrm{Nat}}\right)^{2}}{\left(|i-j|^{0.15}\right)^{2}}\right],
\end{equation}
which measures the nativeness of the current configuration. It varies between zero and one, where one is the value of the native structure.

\section{Results and Discussion}
\begin{figure}
	\centering
	\includegraphics[width=1\columnwidth,height=0.25\textheight,keepaspectratio]{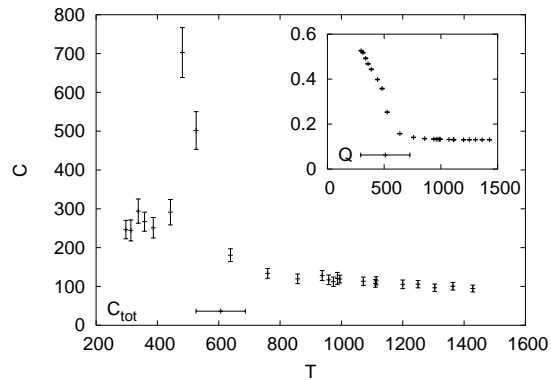}
	\caption{Specific heat vs.~T. of an unbiased parallel tempering run. The sharp specific heat peak at $T_1=481$ is correlated with the helix-coil transition}
	\label{fig:C-tot-w-Q-inset-noGo}
\end{figure}

\begin{figure}
 \centering
 \includegraphics[width=1\columnwidth, height=0.25\textheight, keepaspectratio]{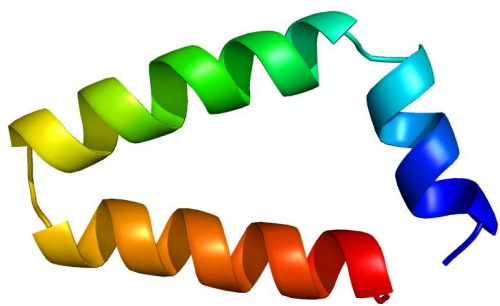}
 \caption{Minimum energy configuration from the unbiased parallel tempering run. The N-terminal helix, has the wrong orientation and the rmsd over
all residues is therefore with 8.8 \AA\  large. The all-atom rmsd for residues 16--46 is 3.2 \AA.}
 \label{fig:min-eaa-kGo-0.0}
\end{figure}

\begin{figure}
 \centering
 \includegraphics[width=1\columnwidth,height=0.25\textheight,keepaspectratio]{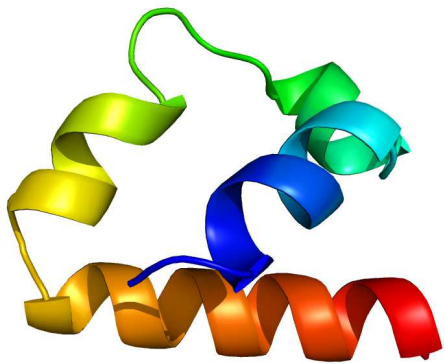}
 \caption{Most native like configuration from the unbiased parallel tempering run. The orientation of the N-terminal helix is correct leading to an all-atom rmsd of 3.4 \AA\ over all residues.}
 \label{fig:max-native-kGo-0.0}
\end{figure}

We start by presenting our results for a regular parallel tempering simulation 
without any Go potential $(k_\mathrm{Go}=0)$. Our simulation used 24 replicas
with temperatures varied between 297 and 1429K. Starting from a stretched
configuration we performed 100,000 sweeps. \Fref{fig:C-tot-w-Q-inset-noGo}
displays the specific heat as
a function of temperature. The temperature set was optimized following
the suggestions by Trebst and Hansmann \cite{Trebst05}
We observe a steep peak at a temperature of $T_1=481$ K
followed by a more broader saddle at a second and lower temperature $T_2=338$ K. 
The two transitions are also visible in our order parameter $Q$ that is displayed
in the inset. The steep increase at the higher temperature $T_1$ is correlated with
a helix-coil transition at this temperature (data not shown), i.e., the formation of short-range
contacts, while the second transition at lower temperature $T_2$ marks formation of 
long range contacts. \Fref{fig:min-eaa-kGo-0.0} 
displays the configuration 
with lowest energy obtained in the simulation. It has an  all-atom rmsd of 3.2 \AA\ for 
residue 16--46. The N-terminal helix, however, has the wrong orientation and the rmsd over
all residues is therefore with 8.8 \AA\  large. The 
configuration with highest value of $Q$ is displayed in \fref{fig:max-native-kGo-0.0}.
Here, the orientation of the N-terminal helix is correct leading to an all-atom rmsd of 
3.4 \AA\ (over all residues) and a solvent accessible surface area of 3680 \AA$^2$ that is smaller than 
the one (4340 \AA$^2$) for the minimal energy configuration of \fref{fig:min-eaa-kGo-0.0}.
 However, the energy of this configuration is with $E = - 567.2$ kcal/mol almost $50$ kcal/mol higher
than that of the minimal energy configuration ($E=-614.6$ kcal/mol). This is because the
ECEPP force field over-emphasizes helix formation. For protein A this leads to formation of three
helices  that are more elongated than  observed in the PDB structure and therefore are too stiff to
arrange themselves into the correct configuration. Consequently, the higher
energy of the configuration with maximal order parameter $Q$ is due to the intra-molecular
energy term $E_\mathrm{ECEPP}$ ($-378.0$ kcal/mol vs. $-431.5$ kcal/mol) while the
solvation energy $E_\mathrm{solv}$ is slightly lower ($-189.2$ kcal/mol vs. $-183.1$ kcal/mol. 
From our result it is not clear whether the global minimum  energy configuration  would
be native-like and was just not found in the simulation, or whether it differs for this
force field from the native structure of \fref{fig:Protein-A}. In either case this indicates problems with
our energy function that limits its use in protein simulations.   

\begin{figure}
 \centering
 \includegraphics[width=1\columnwidth,height=0.25\textheight, keepaspectratio]{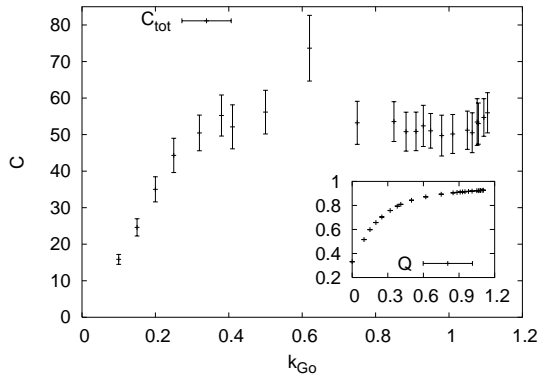}
 \caption{Specific heat C vs.~Go parameter $k_\mathrm{Go}$. There is no apparent specific heat peak. The inset shows a smooth increase in Q. The structures become increasingly native-like as $k_\mathrm{Go}$ increases.}
 \label{fig:C-tot-w-Q-inset-kaa-0.0}
\end{figure}

\begin{figure}
 \centering
 \includegraphics[width=1\columnwidth,height=0.25\textheight, keepaspectratio]{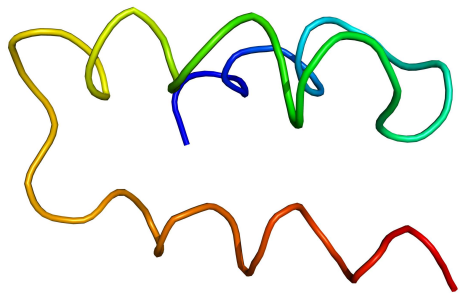}
 \caption{Minimum energy configuration from a parallel tempering with Go energy only. The Go energy does not distinguish between the native structure and its mirror image. In this run we obtained the mirror image of the native structure.}
 \label{fig:min-e-kaa-0}
\end{figure}

 The situation is different in simulations with a Go-energy function. Here, it is by definition of
 the energy ensured that the global minimum configuration is the native structure (or its mirror
 configuration).  This can be seen in \fref{fig:C-tot-w-Q-inset-kaa-0.0} which displays 
 the results from a simulation with only the Go-term of \eref{go-term}. 
The replicas differ here in the value of the Go-parameter $k_\mathrm{Go}$, i.e., the true inverse temperature in the system
 is $\beta k_\mathrm{Go}$ (with $\beta$ the inverse temperature corresponding to $T=300$ K ). Shown is again the specific heat and in the inset our order parameter Q. The system does not seem to have any transition. The Order parameter is increasing monotonously. No pronounced peak is observed in the specific heat. By construction of the energy function the lowest energy  configuration is also the one with the largest $Q$ value and shown in \fref{fig:min-e-kaa-0}. 
Note that this structure
 is actually a mirror configuration and therefore the rmsd is with 8 \AA\  larger than one would expect
 from visual inspection.
 
\begin{figure}
 \centering
 \includegraphics[width=1\columnwidth,height=0.25\textheight,keepaspectratio]{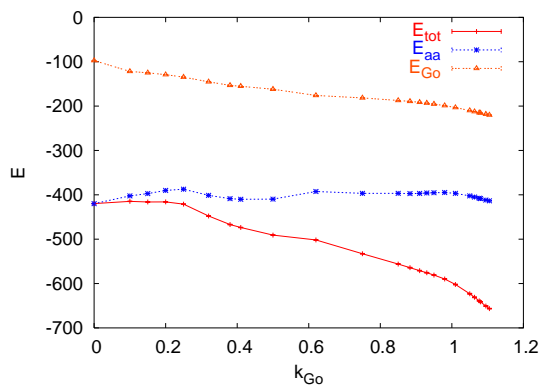}
 \caption{Total, all-atom, and Go energy vs. $k_\mathrm{Go}$ at constant temperature T=300K. At $k_\mathrm{Go}\approx 0.2$ the Go energy starts to dominate the behaviour of the total energy.}
 \label{fig:E-avg-kaa-1.0-Taa-300}
\end{figure}

 In the following we study now how the bias introduced by a Go-term affects the outcome of
 an all-atom simulations. For this purpose we study our protein at temperature $T =300 $K
 just below the folding temperature $T_2$, varying the strength with that the Go-term contributes
 to the total energy of the system over the ladder of replicas. 
 \Fref{fig:E-avg-kaa-1.0-Taa-300} shows the various energy terms
 as a function of the coupling strength $k_\mathrm{Go}$  of the Go-term. 
As expected, the Go-energy decreases with  increasing strength of coupling. However, the all-atom energy stays constant, i.e. does not
 change with the  introduction of the additional Go-term. The superposition of the two energy
 terms leads to a total energy $E_\mathrm{tot}$ that sharply decreases for $k_\mathrm{Go} \ge 0.2$. 
  Hence, for a ``critical''  $k_\mathrm{Go}$ the contribution from the Go-term starts dominating the system.
  We therefore conjecture that $k_\mathrm{Go}= 0.2$ is  the optimal value for coupling of the two energy terms. For a lower value, the influence
 of the Go term is too weak to be effective, while for a larger value the system behaves as a
 Go-model.
 
\begin{figure}
 \centering
 \includegraphics[width=1\columnwidth,height=0.25\textheight,keepaspectratio]{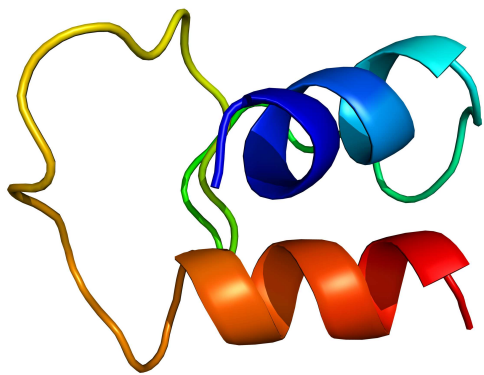}
 \caption{Minimum energy configuration from a biased parallel tempering run with $k_\mathrm{Go}=0.2$.}
 \label{fig:min-e-kGo-0.2}
\end{figure}

 Fixing now the $k_\mathrm{Go} = 0.2$ we perform again a parallel tempering simulation in temperature.
 The resulting lowest energy configuration is shown in \fref{fig:min-e-kGo-0.2}.
 It has an 
 all-atom rmsd of 4.5 \AA\  over all residues (compared to 8.8 \AA\  for the case without a coupled 
 Go-term). When comparing the all-atom energies, we find that the value for this configuration
 $E_\mathrm{aa} = -580.4$ kcal/mol is higher than that of the free case ($E_\mathrm{aa} = -614$ kcal/mol). Hence,
 it is not so that the additional Go-term solely smoothens the energy landscape and increases in this
 way the chances of finding a native-like configurations as the true global minimum. Rather, we conjecture that for Protein A the global minimum in our all-atom force field is not the native structure. Only by adding the Go-term is the energy landscape deformed in a way that the native structure (being a sub-optimal competing local minimum in the all-atom energy) becomes the global minimum in the total
 energy.

   \section{Conclusions}
   We have performed simulations of the 46 amino-acid long segment of Protein A. Simulating
   the protein with a ``physical'' all-atom force field we find low-energy configurations that are similar
   to the native structure but the global minimum configuration differs significantly (by $\approx 8$ \AA)
   from this.  Addition of a Go-term leads in the simulation to a global minimum (in the combined energy) 
   that is close to the native one. However, its all-atom energy is higher than the one found for the
   global minimum found in a simulation relying only on an all-atom force field. We conclude that the
   Go-term deforms the energy landscape in a way that the native structure becomes the global minimum
   in the combined energy but that it is not the one for the all-atom force field. As the introduction of the Go-term does not account to mere smoothening of the energy landscape but to a larger deformation of the energy landscape  it is not a suitable tool for the faster thermalization of
   all-atom simulations.

\ack
U.H. acknowledges support by a research grant (CHE-0313618) of the National
Science Foundation (USA). The simulations were done on the Cray XD1 and the JUMP supercomputer of the John von Neumann Institute for Computing at the Research Center J\"{u}lich.

\section*{References}
\bibliographystyle{unsrt}
\bibliography{go}

\begin{thebibliography}{10}

\bibitem{Anfinsen1973}
Christian~B. Anfinsen.
\newblock Principles that govern the folding of proteins.
\newblock {\em Science}, 181:223--230, 1973.

\bibitem{Onuchic1997}
Jose~Nelson Onuchic, Zaida Luthey-Schulten, and Peter~G. Wolynes.
\newblock {\em Annu. Rev. Phys. Chem.}, 48:545, 1997.

\bibitem{Abe81}
H.~Abe and N.~Go.
\newblock Noninteracting local-structure model of folding and unfolding
  transition in globular proteins. ii. application to two-dimensional lattice
  proteins.
\newblock {\em Biopolymers}, 20(5):1013 -- 1031, 1981.

\bibitem{Go1983}
Nobuhiro Go.
\newblock Theoretical studies of protein folding.
\newblock {\em Ann. Rev. Biophys. Bioeng}, 12:183--210, 1983.

\bibitem{Takeda1999}
Shoji Takada.
\newblock Go-ing for the prediction of protein folding mechanisms.
\newblock {\em Proc. Natl. Acad. Sci. USA}, 96:11698--11700, 1999.

\bibitem{Hansmann2004}
Ulrich H.~E. Hansmann.
\newblock Generalized-ensemble simulations of the human parathyroid hormone
  fragment pth(1-34).
\newblock {\em J. Chem Phys}, 120(1):417--422, 2004.

\bibitem{Hansmann2003}
Chai-Yu Lin, Chin-Kun Hu, and Ulrich H.~E. Hansmann.
\newblock Parallel tempering simulations of {HP}-36.
\newblock {\em Proteins}, 52(3):436--45, 2003.

\bibitem{Trebst05}
S.~Trebst, M.~Troyer, and U.~H.~E. Hansmann.
\newblock Optimized parallel tempering simulations of proteins.
\newblock {\em J. Chem. Phys.}, 124(17):174903, 2006.

\bibitem{Pogorelov2004}
Taras~V. Pogorelov and Zaida Luthey-Schulten.
\newblock Variations in the fast folding rates of the l-repressor: A hybrid
  molecular dynamics study.
\newblock {\em Biophys. J.}, 87:207--214, 2004.

\bibitem{ECEPP2}
Manfred~J. Sippl, George N{\'{e}}mthy, and Harold~A. Scheraga.
\newblock Intermolecular potentials from crystal data. 6. determination of
  empirical potentials for o-h***o=c hydrogen bonds from packing
  conflguratlons.
\newblock {\em J. Phys. Chem.}, 88:6231--6233, 1984.

\bibitem{ECEPP3}
G.~Nemethy, K.~D. Gibson, K.~A. Palmer, C.~N. Yoon, G.~Paterlini, A.~Zagari,
  S.~Rumsey, and H.~A. Scheraga.
\newblock Energy parameters in polypeptides. 10. improved geometrical
  parameters and nonbonded interactions for use in the {ECEPP/3} algorithm,
  with application to proline-containing peptides.
\newblock {\em Journal of Physical Chemistry}, 96(15):6472 -- 6484, 1992.

\bibitem{Eisenmenger2001}
Frank Eisenmenger, Ulrich H.~E. Hansmann, Shura Hayryan, and Chin-Kun Hu.
\newblock {[SMMP] A modern package for simulation of proteins}.
\newblock {\em Computer Physics Communications}, 138:192--212, aug 2001.

\bibitem{Eisenmenger2006}
Frank. Eisenmenger, Ulrich.~H.~E. Hansmann, Shura Hayryan, and Chin-Kun Hu.
\newblock An enhanced version of {SMMP}---an open-source software package for
  simulation of proteins.
\newblock {\em Computer Physics Communications}, 174:422--429, 2006.

\bibitem{OONS}
Tatsuo Ooi, Motohisa Oobatake, George Nemethy, and Harold~A. Scheraga.
\newblock Accessible surface areas as a measure of the thermodynamic parameters
  of hydration of peptides.
\newblock 84:3086--3090, 1987.

\bibitem{PT1}
Koji Hukushima and Koji Nemoto.
\newblock Exchange {M}onte {C}arlo method and application to spin glass
  simulations.
\newblock {\em Phys. Soc. (Jap)}, 65:1604 -- 1608, 1996.

\bibitem{PT2}
C.~J. Geyer and E.~A. Thompson.
\newblock Annealing {M}arkov chain {M}onte {C}arlo with applications to
  ancestral inference.
\newblock {\em J. Am. Stat. Assn.}, 90:909, 1995.

\bibitem{Hansmann1997}
Ulrich H.~E. Hansmann.
\newblock Parallel tempering algorithm for conformational studies of biological
  molecules.
\newblock {\em Chem. Phys. Lett.}, 281(1--3):140--150, 12 1997.

\bibitem{Friedrichs1989}
Mark~S. Friedrichs and Peter~G. Wolynes.
\newblock Towards protein tertiary structure recognition by means of
  associative memory hamiltonians.
\newblock {\em Science}, 246(4928):371--373, 1989.

\bibitem{Hardin1999}
Corey Hardin, Zaida Luthey-Schulten, and Peter~G. Wolynes.
\newblock Backbone dynamics, fast folding, and secondary structure formation in
  helical proteins and peptides.
\newblock {\em Proteins}, 34(3):281--294, 1999.

\end{thebibliography}
\end{document}